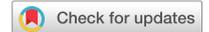

# SCIENTIFIC DATA

**OPEN**

**DATA DESCRIPTOR**

# High-resolution monthly precipitation and temperature time series from 2006 to 2100

Dirk Nikolaus Karger✉, Dirk R. Schmatz, Gabriel Dettling & Niklaus E. Zimmermann

Predicting future climatic conditions at high spatial resolution is essential for many applications and impact studies in science. Here, we present monthly time series data on precipitation, minimum- and maximum temperature for four downscaled global circulation models. We used model output statistics in combination with mechanistic downscaling (the CHELSA algorithm) to calculate mean monthly maximum and minimum temperatures, as well as monthly precipitation at ~5 km spatial resolution globally for the years 2006–2100. We validated the performance of the downscaling algorithm by comparing model output with the observed climate of the historical period 1950–1969.

## Background & Summary

High-resolution future climate projections are essential for many scientific applications ranging from climate change impact studies, environmental planning, and ecological analysis and modelling. There is, however, a relatively large-scale gap between the spatial resolution at which global circulation models (GCMs) are calculated, and the resolution at which impact studies are conducted. While many studies in ecology and environmental sciences are conducted at a relatively fine spatial resolution of just a few kilometres, GCMs represent climatic variation at spatial resolutions of 0.5°–1° (ca. 50–100 km at the equator). Such a coarse resolution is usually not capable of capturing orographic precipitation in complex terrain[1–3]. Although global circulation and weather models such as WRF-ARF[4], or ICON[5,6], for example, can be run at high horizontal resolutions close to 1 km, they are still heavily constrained by computational limits[7]. Currently, global kilometre-scale models only achieve a simulation throughput of 0.043 SYPD (simulated years per day)[8], which amounts to an 25 x shortfall compared to what would be computationally efficient simulations of 1 SYPD[7,9]. Even with the largest supercomputers and state-of-the-art climate models, as well as large financial investments, such a shortfall can currently only be reduced by a factor of 20 (ref. [10]).

Although achieving 1 km resolutions in numerical climate modelling is important for quantifying effects such as deep convection or surface drag[10], impact studies usually rely only on a much simpler set of climate variables compared to what numerical climate models can provide. Therefore, impact studies do not require a complete representation of all climate processes at high resolution. In ecological studies for example, precipitation together with minimum and maximum temperatures are often used to analyze occurrences of species[11]. Also, it is common to characterize species ranges by their climatic envelopes using species distribution models (SDMs) and a relatively small set of climatic predictors derived from monthly minimum and maximum temperature, as well as precipitation[12,13].

For many scientific applications, the representation of the temporal and spatial variability of temperature and precipitation is extremely important[14]. The gap between these spatial scales is often bridged by applying a delta change method[15,16] to current-time climate data that is available at high spatial resolution of ca. 1–20 km e.g. from CHELSA[17], WorldClim[18], CRU[19], GPCC[20], CHIRPS[21], or PRISM[22]. Such datasets exist usually only for climatological means calculated for specific time periods, but time series that allow for a more dynamic representation of the climate system are still missing at high resolutions of ca. 1 km. Up to now, only downscaled mean climatological data at high spatial resolution has been available for the future (e.g. ref. [16]), yet high resolution (<10 km) global time series are still lacking.

Here, we present four downscaled global circulation models (GCMs) from the Coupled Model Intercomparison Project Phase 5 (CMIP5[23]) gridded monthly time series for the years 2006–2100 with a spatial resolution of 0.049° resolution (approximately 5 km at the equator). The downscaling algorithm is based on

Dynamic Macroecology, Swiss Federal Institute for Forest, Snow and Landscape Research WSL, Zürcherstrasse 111, 8903, Birmensdorf, Switzerland. ✉e-mail: dirk.karger@wsl.ch

                                                                                                     1

the CHELSA algorithm[17], which provides a more accurate representation of temperature and precipitation in highly complex terrain. Data based on the CHELSA algorithm[17] have already been used to infer e.g. ecological niches[24], soil nutrients[25,26], or to assess climate change impacts of forests[27], insect pests[28], and biodiversity[29]. Mean monthly maximum daily temperatures and mean monthly minimum daily temperatures have been downscaled using the delta change method based on the high-spatial-resolution data from CHELSA V.1.2. Monthly precipitation sums have been downscaled by applying the CHELSA algorithm directly on bias corrected GCM data. The CHELSA algorithm allows for representing the effects from changing wind patterns and boundary layer conditions in the process of downscaling, and therefore allows for a better estimation of the km-scale changes in precipitation patterns under future climate projections.

## Methods

**Selection of global circulation models (GCMs).** Projected future climate variables were taken from four global circulation models (GCMs) driven by two scenarios of representative concentration pathways (RCPs) in a factorial manner. The four selected models originate from the CMIP5 collection of model runs used in IPCC's 5th Assessment Report[30] (IPCC 2013). GCMs are, however, often based on similar code which consequently results in similar output[31,32]. We therefore chose models that show a low amount of interdependence to allow for a good representation of uncertainty among available climate projections. Model selection was performed to reduce model interdependence in ensembles (see ref.[32]).

The four models from which data were taken are: CESM1-BGC run by National Center for Atmospheric Research (NCAR); CMCC-CM run by the Centro Euro-Mediterraneo per I Cambiamenti Climatici (CMCC); MIROC5 run by the University of Tokyo; and ACCESS1-3 run by the Commonwealth Scientific and Industrial Research Organization (CSIRO) and Bureau of Meteorology (BOM), Australia.

**Temperature.** To downscale minimum and maximum temperatures we apply a simple climatologically aided interpolation CAI[33,34] on anomalies derived from the CMIP5 GCMs. Although more sophisticated methods exist to bias-correct and downscale GCM output such as trend-preserving quantile mapping[35], they are usually computationally much more demanding, and therefore hard to implement globally at high resolution. Additionally, precipitation and temperature varies on scales that are much smaller than the grid spacing of the GCM, which means that fine scale spatial variability can be distorted substantially when quantile mapping is used directly for downscaling purposes[36–38]. We therefore applied a monthly climatological aided interpolation based on monthly CHELSA V1.2 time series temperature data spanning the years 1979–2005 $tas_{cur}^{mod}$. Monthly anomalies $\Delta R_m^{mod}$ are calculated based on the respective CMIP5 model output using the difference between today $tas_{cur}^{mod}$ and future $tas_{fut}^{mod}$ on the resolution of the respective GCM. As the timeframe of the CHELSA data does not fully cover the historical period of the CMIP5 runs, we used the years 1979–2005 from historical runs, as reference period for the GCM climatologies. To achieve a gap-free anomaly grid surface, we interpolated $\Delta R_m^{mod}$ using a multilevel B-spline interpolation[39] with 14 error levels optimized using B-Spline refinement[39] to the 5 km grid resolution. The multilevel B-spline approximation applies a B-spline approximation to $\Delta R_m^{mod}$ *starting* with the coarsest lattice $\phi_0$ from a set of control lattices $\phi_0, \phi_1, ..., \phi_n$ with $n = 14$ that have been generated using optimized B-Spine refinement[39]. The resulting B-spline function $f_0(\Delta R_m^{mod})$ gives the first approximation of $\Delta R_m^{mod} \cdot f_0(\Delta R_m^{mod})$ and leaves a deviation between $\Delta^1 R_m^{mod} c = \Delta R_m^{mod} - f_0(x_c, y_c)$ at each location $(x_c, y_c, \Delta R_m^{mod} c)$. Then the next control lattice $\phi_1$ is used to approximate $f_1(\Delta^1 \Delta R_m^{mod} c)$. Approximation is then repeated on the sum of $f_0 + f_1 = \Delta R_m^{mod} - f_0(x_c, y_c) - f_1(x_c, y_c)$ at each point $(x_c, y_c, \Delta R_m^{mod} c)$ $n$ times resulting in the gap free interpolated bias surface $\Delta R_m^{int}$. The bias correction surface $\Delta R_m^{int}$ is then added to the observed temperatures at high-resolution $tas_{cur}^{mod}$ to get the anomaly corrected monthly temperatures using:

$$tas = tas_{cur}^{obs} + R_m^{int} \qquad (1)$$

**Precipitation.** Elevation is the main topo-climatic driver of vertical precipitation gradients, but the relation between elevation and precipitation can be idiosyncratic[40–46]. In convective regimes of the tropics, precipitation amounts commonly increase up to the condensation level at about 1000–1500 m above the ground surface, and the exponentially decreasing air moisture content in the mid- to upper troposphere results in a corresponding drying above the condensation level of tropical convection cluster systems (non-linear precipitation lapse rates)[47]. Likewise, negative lapse rates typically occur in the extremely dry polar climates. At mid-latitudes and in the subtropics, the frequent or even prevalent advection of moisture bearing air to high altitudes generally results in increasing precipitation with increasing elevation. Consequently, the summits of high mountain ranges such as the Alps[48] by tendency exhibit high rainfall, and the associated lapse rates for precipitation are quasi-linear[49]. Reduced precipitation at lower elevations is due, firstly, to the evaporation of raindrops when falling through non-saturated, lower-air levels. Secondly, the vertical precipitation gradient in high mountain ranges is often increased due to the diurnal formation of autochthonous upslope breezes. This upward flow of air intensifies cloud and precipitation formation in upper slope positions whilst the subsiding branch of these autochthonous local circulation systems along the valley axis leads to cloud dissolution and a corresponding reduction of precipitation rates in the valley bottoms. We approximated such orographic precipitation effects using the semi-mechanistic CHELSA downscaling algorithm[17] as explained in detail below.

We applied simple monthly change factor bias corrections to remove the inherent bias of the GCM precipitation $pr^{mod}$ before the downscaling by using the historical period from 1979–2005 from CHELSA $pr^{obs}$ as a reference period. The monthly bias $R_m$ is then calculated using:

$$R_m = \frac{pr^{obs} + c}{pr^{mod} + c}$$

(2)

The bias correction surface $R_m$ was then multiplied with the GCM values to get the corrected precipitation values. The constant $c$ was set to 1 to avoid divisions by zero.

**Wind effect correction.** Orographic effects are among the most important drivers of precipitation[44,50–53]. Such orographic effects[54] caused by wind fields are the most common influence of small-scale variations in precipitation[48,52,54–56]. To include the effect of orographic barriers, we used a wind field index[17,41,57] to account for the expected higher precipitation at windward compared to leeward sites of an orographic barrier. We used the monthly u-wind and v-wind components from the respective GCM at surface level as representatives of underlying wind components. These two wind vectors were interpolated to the high (~5 km) grid resolution using a multilevel B-spline interpolation similar to the one used to interpolate $\Delta R_m^{int}$. As the calculation of a windward-leeward index (hereafter: wind effect) requires a projected coordinate system, both wind components were projected to a world Mercator projection and then combined to a directional grid. The wind effect $H$ with windward component $H_W$ and the leeward component $H_L$ was then calculated using:

$$H_{W,L} = \begin{cases} \dfrac{\sum_{i=1}^{n} \frac{1}{d_{WHi}} tan^{-1}\left(\frac{d_{WZi}}{d_{WHi}^{0.5}}\right)}{\sum_{i=1}^{n} \frac{1}{d_{LHi}}} + \dfrac{\sum_{i=1}^{n} \frac{1}{d_{LHi}} tan^{-1}\left(\frac{d_{LZi}}{d_{LHi}^{0.5}}\right)}{\sum_{i=1}^{n} \frac{1}{d_{LHi}}}, & d_{LHi} < 0 \\[20pt] \dfrac{\sum_{i=1}^{n} \frac{1}{d_{WHi}} tan^{-1}\left(\frac{d_{LZi}}{d_{WHi}^{0.5}}\right)}{\sum_{i=1}^{n} \frac{1}{d_{LHi}}}, & d_{LHi} \geq 0 \end{cases}$$

(3)

where $d_{WHi}$ and $d_{LHi}$ refer to the horizontal search ranges (here 75 km) in windward and leeward direction and $d_{WZi}$ and $d_{LZi}$ are the corresponding vertical distances compared with the considered grid cell. The second summand in the equation for $H_{W,L}$ where $d_{LHi} < 0$ accounts for the leeward impact of previously traversed mountain chains. The horizontal distances in the equation for $H_{W,L}$ where $d_{LHi} \geq 0$ lead to a longer-distance impact of leeward rain shadow. The final wind-effect parameter, which is assumed to be related to the interaction of the large-scale wind field and the local-scale precipitation characteristics, is calculated as $H = H_{W,L} \to d_{LHi} < 0 * H_{W,L} \to d_{LHi} \geq 0$ and generally takes values between 0.7 for leeward and 1.3 for windward positions. Both equations were applied to each grid cell at the 30 arc-sec resolution in a World Mercator projection.

**Valley exposition correction.** Although the wind effect algorithm can distinguish between windward and leeward sites of an orographic barrier, it cannot distinguish extremely isolated valleys inside highly-elevated mountain areas[58]. Such valleys are situated in areas where the wet air masses flow over an orographic barrier and are prevented from flowing into deep valleys. These effects are however mainly confined to large mountain ranges, and are not as prominent in small- to intermediate-sized mountain ranges[59]. To account for these effects, we used a variant of Eq. (3) with a linear search distance of 300 km in steps of 5° from 0° to 355° circular angles for each grid cell. The calculated leeward index was then scaled towards higher elevations using:

$$E = H_L^{\frac{elev}{c}}$$

(4)

which rescales the strength of the exposition index relative to elevation ($elev$) from GMTED2010, and assigns valleys located in high mountain areas with larger wind isolations ($E$) than valleys located at low elevations. The correction constant $c$ was set to 9000 m to include all possible elevations of the DEM similar to the CHELSA V1.2 algorithm. The constant has been set to 9000 m as values of $elev > c$ could lead to a reverse relationship between $elev$ and $H_L$. The valley exposition index was calculated at 30 sec and was then resampled to the 0.044915° grid resolution using the mean function.

**Boundary layer correction.** Orographic precipitation effects are less pronounced just above the surface, as well as in the free atmosphere above the planetary boundary layer[22,60,61]. The highest impact of orography is considered just at the boundary layer height where the airflow interacts with the terrain. We used the boundary layer height $B$ from the GCM as indicator of the pressure level that exerts the highest contribution to the wind effect. The boundary layer height has been interpolated to the 0.044915° grid resolution using a B-spline. The wind effect grid $H$ containing the windward ($H_W$) and leeward ($H_L$) index values was then proportionally distributed to all grid cells falling within a respective GCM grid cell using:

$$H_{WB} = \frac{H_W}{1 - \left(\frac{|d| - d_{max}}{c}\right)}$$

(5)

| Attribute | Value |
|---|---|
| Resolution (decimal degrees): | 0.0449147848 |
| West extent (minimum X-coordinate, longitude): | −180.0184296146 |
| South extent (minimum Y-coordinate, latitude): | −90.0184296146 |
| East extent (maximum X-coordinate, longitude): | 179.9286557726 |
| North extent (maximum Y-coordinate, latitude): | 83.9365319158 |
| Rows: | 3874 |
| Columns: | 8015 |
| Projection: | Geographic Coordinate System |
| EPSG: | 4326, WGS84 |
| Proj4: | +proj=longlat +datum=WGS84 +no_defs |

**Table 1.** Grid extent and resolution of the netCDF files.

$$H_{LB} = \frac{H_L}{1 - \left( \frac{|d| - d_{max}}{c} \right)} \tag{6}$$

with:

$$d = elev - B \tag{7}$$

With $d$ being the distance between a grid cell and the boundary layer height $B$, $d_{max}$ being the maximum distance between the boundary layer height $B$ and all grid cells at the 0.044915° grid resolution falling within a respective GCM grid cell, $c$ being a constant of 9000 m, and $elev$ being the respective elevation from GMTED2010. with:

$$B = B_{GCM} + elev_{GCM} \tag{8}$$

$B$ is the height of the monthly means of daily mean boundary layer from the GCM, and $elev_{GCM}$ is the elevation of the GCM grid cell. The archived precipitation values at the 0.5° grid resolution are then downscaled to the final resolution of 0.044915° using:

$$pr = \frac{H}{\overline{H}} * pr_m \tag{9}$$

where $\overline{H}$ is the mean wind effect at the GCM resolution, $pr_m$ is the bias corrected precipitation rate $pr^{mod} * R_m$, and $H$ is the boundary layer corrected wind effect surface. By using a linear relationship, we force the downscaled data, e.g. precipitation, of all 0.044915° cells within the range of a grid cell to match that of the GCM precipitation.

## Data Records

The dataset[62] is available on EnviDat (https://doi.org/10.16904/envidat.124) and consists of geographical rasters in netCDF4 format at a 0.0449147848° spatial resolution for a total of three variables, four GCMs, and two RCPs. All variables are available separately for each month starting Jan. 1st 1850, and ending Dec. 31st 2100. Parameters of the separate raster files are listed in Table 1.

## Technical Validation

**Performance of the downscaling algorithm.** We used station data of the Global Historical Climatology Network-Daily (GHCN-D) database[35] to validate the model predictions. We used the observations from January 1950 to December 1969 as the validation period. This period is available in the different CMIP5 model runs, as well as in the observational database. For each station, we calculated monthly mean temperature and precipitation but discarded stations with less than 25 daily values per month. Although CMIP5 models are not designed to accurately reflect the weather conditions during these times, the comparison of GHCN station data to both the original and the downscaled GCM data still offers a way to evaluate the overall performance of the downscaling algorithm.

The downscaled CMIP5 precipitation data clearly shows a much lower root mean square error (RMSE), as well as a lower mean absolute error (MAE) when compared to the CMIP5 data at the original resolution (Fig. 1). At the same time the correlation coefficient of the downscaled data increases clearly for all models compared to the original GCMs (Fig. 1) for all three variables (tasmax, tasmin, and pr).

**Maximum temperatures.** The downscaled CHELSA$_{CMIP5}$ts maximum temperature data shows a lower deviation, as well as a higher correlation compared to the original CMIP5 model data equally well for both rcps (Fig. 2). There is no significant difference in the correlation coefficient between the original and the downscaled data (Fig. S1). For both rcp's, predicted values from CHELSA$_{CMIP5}$ts are closer to the observed values than the original CMIP5 data, although there is still not a perfect fit. The remaining difference might be explained by the differences in elevation of a GHCN station to that of a grid cell, general errors in the station measurements, or the fact that in GCMs in general are not able to accurately predict the weather, but rather the overall variance in the climate system. However, after downscaling, the GCM predictions are still closer to the observed values than those of the original GCM, indicating that the downscaling algorithm can offer some improvement over the use of the original data.

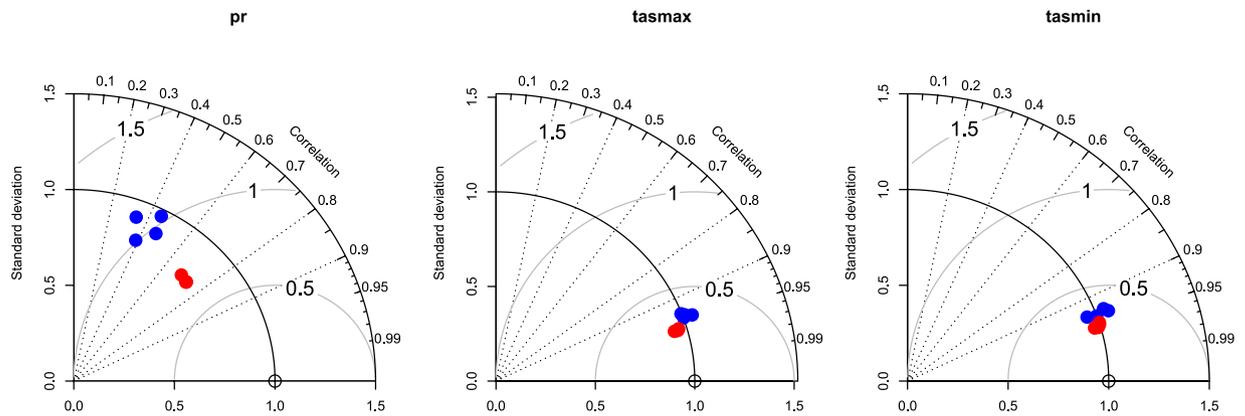

**Fig. 1** Taylor plots for comparisons between the original GCM data (blue) and CHELSA$_{CMIP5}$ts data (red) for all four GCMs (ACCESS1-3, CESM1-BGC, CMCC-CM, MIROC5). The original data does not perform well with respect to precipitation (ps), but this performance is improved by the downscaling algorithm, although the standard deviation of the downscaled data is slightly higher than that of the original GCMs. For both maximum (tasmax), and minimum (tasmin) temperatures, the original GCM data already shows a good performance, which is further increased by the downscaling to higher resolutions.

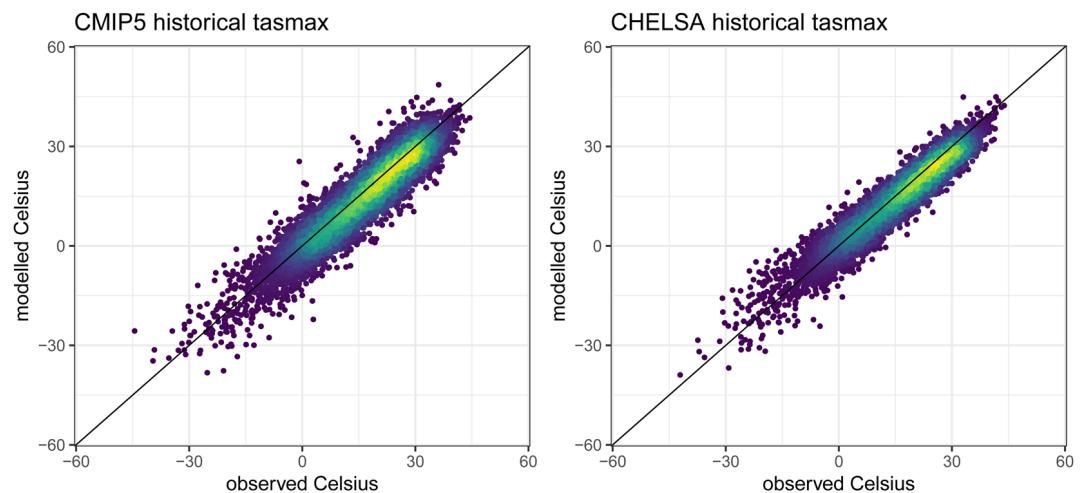

**Fig. 2** Predicted vs observed monthly mean of daily maximum temperatures for the years 1950–1969 from a CMIP5 ensemble including all four used GCMs (ACCESS1-3, CESM1-BGC, CMCC-CM, MIROC5), compared to the downscaled data from CHELSA$_{CMIP5}$ts for the same ensemble. Observed data is based on GHCN daily aggregated to monthly data for stations with sufficient data (>25 days of record per month). Although both ensembles show a high fit compared to observed data, the overall error is reduced by the downscaling algorithm.

Additionally, there is a decrease in the variance of the monthly correlation coefficients between observed and modelled data, which could be interpreted as a higher robustness of the downscaled data compared to the original models (Fig. S1).

To further investigate the performance of the downscaled data with respect to error reduction we compared the different temperature estimates separately for each month.

For maximum temperatures, the downscaling algorithm performance varied for the different GCMs (Fig. S1). For ACCESS1-3, CMCC-CM, and MIROC5 there was a clear improvement in errors after downscaling, while for CESM1-BGC no such improvement was observed. From November to March, correlations with observations of the downscaled data where slightly lower than those of the original GCMs, while for the rest of the year higher correlations of the downscaled data were achieved relative to the correlations of the different GCMs to observations.

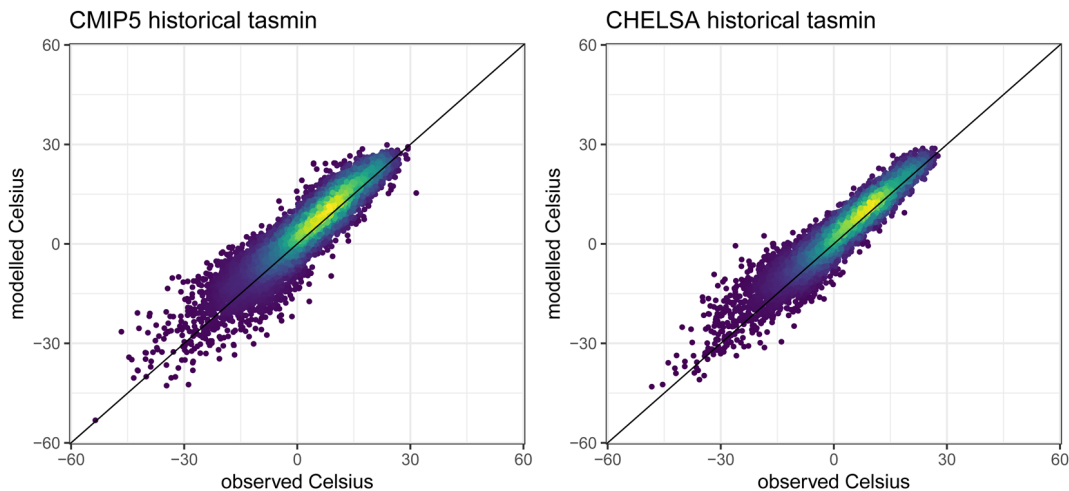

**Fig. 3** Predicted vs observed monthly mean of daily minimum temperatures for the years 1950–1969 from a CMIP5 ensemble including all four used GCMs (ACCESS1-3, CESM1-BGC, CMCC-CM, MIROC5), compared to the downscaled data from CHELSA for the same ensemble. Observed data is based on GHCN daily aggregated to monthly data for stations with sufficient data (>25 days of record per month). Although both ensembles show a high fit compared to observed data, the overall error is reduced by the downscaling algorithm.

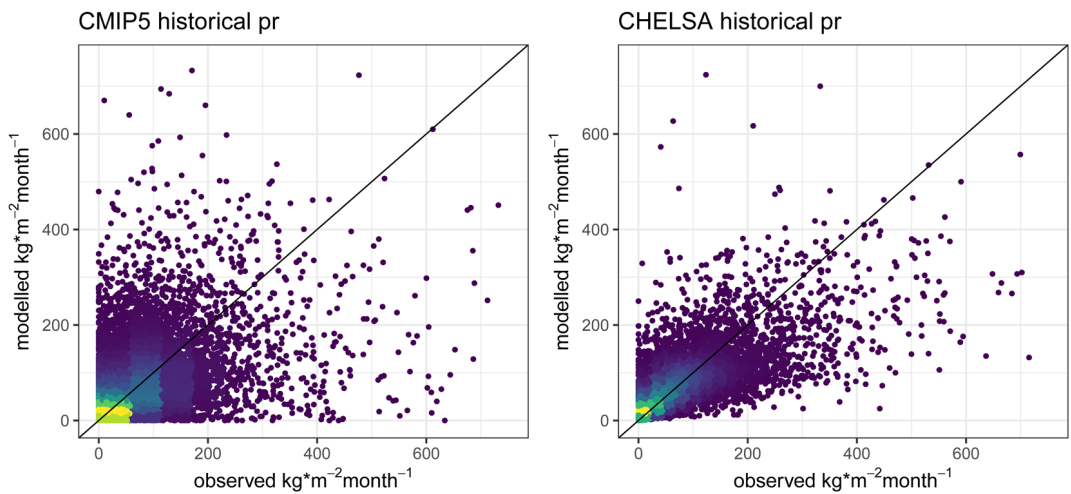

**Fig. 4** Predicted vs observed monthly precipitation sums for 1950–1969 from a CMIP5 ensemble including all four used GCMs (ACCESS1-3, CESM1-BGC, CMCC-CM, MIROC5), compared to the downscaled data from CHELSA for the same ensemble. Observed data is based on GHCN daily aggregated to monthly data for stations with sufficient data (>25 days of record per month). Although the CHELSA algorithm improves the error in the estimated precipitation, it still underestimates precipitation to some degree.

**Minimum temperatures.** For minimum temperatures, a similar pattern emerges as for maximum temperatures, although the improvement after downscaling was only marginal compared to that of maximum temperatures. The variance is improved after downscaling, while the overall correlation is preserved (Fig. 1). This pattern is similar independent of which rcp is used as a basis for the downscaling (Fig. 3). The lower temperatures still show, however, a strong variance, which could be due to the fact that we do not consider cold air flow within the downscaling algorithm.

Still ACCESS1-3, CMCC-CM, and MIROC5 showed some improvement in MAE and RMSE, as well as with their correlation. April to October showed higher correlations again than November to April in which even a decrease in correlation values was observed for CESM1-BGC, CMCC-CM, and MIROC5 (Fig. S2).

**Precipitation.** Improvements of the model algorithm for precipitation compared to the original input data are already visible from the comparison to observations (Fig. 1). However, although the overall estimation of precipitation is improved, the standard deviation is slightly increased. In particular, the algorithm has difficulties to estimate the higher end of the observed precipitation amounts, underestimating high precipitation (Fig. 4). This is a common problem in many climatologies, which can only be overcome by a bias correction incorporating the general gauge undercatch into climate downscaling algorithms[63].

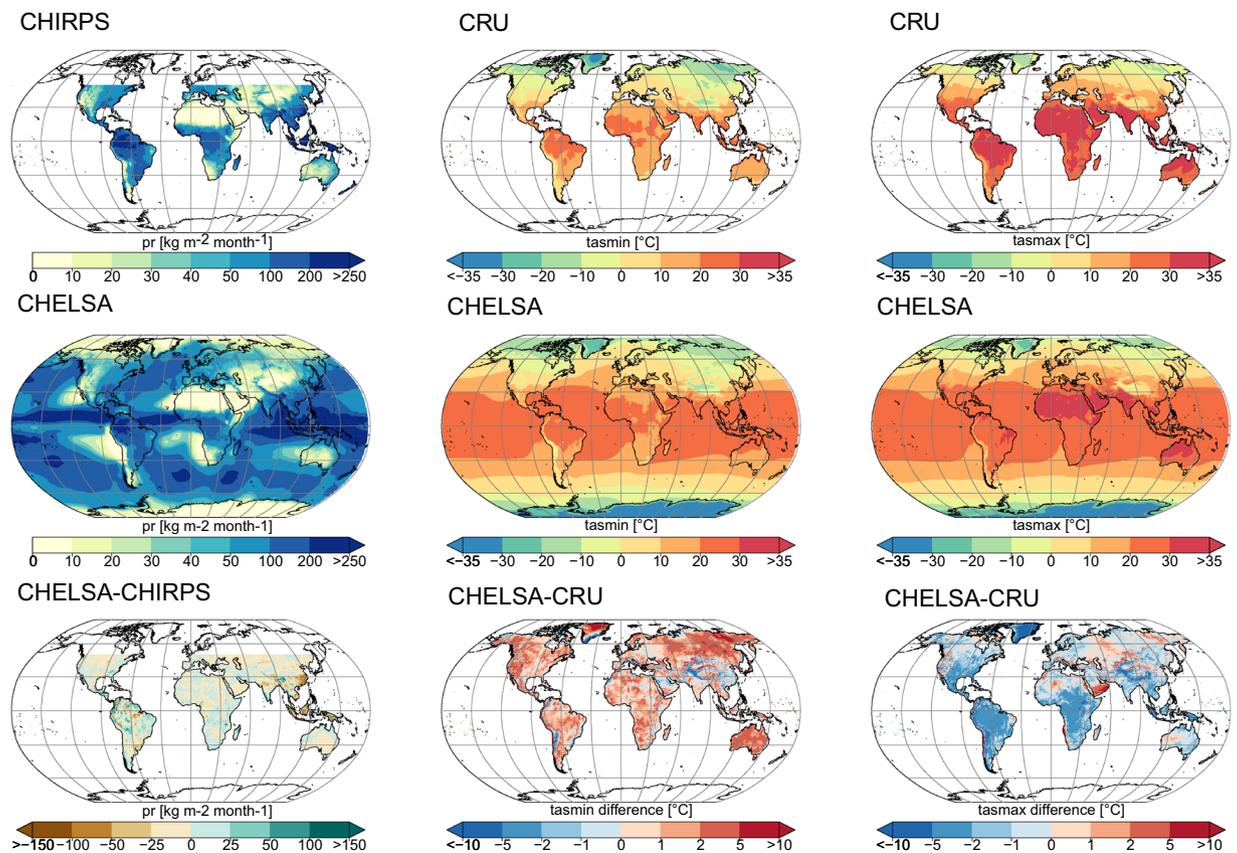

**Fig. 5** Geographical comparison of precipitation (pr), maximum temperatures (tasmax), and minimum temperatures (tasmin) with two different observational datasets (CHIRPS[65] and CRU[64]). All data shown are means over the years 2006–2016. The CHELSA_CMIP5 ts (CHELSA) data consist of an ensemble from ACCESS1-3, CESM1-BGCk, CMCC-CM, and MIROC5 and both rcp's (rcp 4.5 and rcp 8.5).

For precipitation, all four GCMs showed a relative larger bias in the months of January, February, and March when compared to observations than did the downscaled patterns for these months (Fig. S3). After downscaling and bias correction, the errors are smaller overall, are more evenly distributed throughout the year, and correlations increase (Fig. S3).

**Geographical biases.** We use an ensemble of all downscaled CMIP5 GCMs and rcp's by taking their mean and compare it to two observational datasets. For tasmax and tasmin we used the CRU TS v4.01[64] and for precipitation CHIRPS[65] data averaged over the period of 2006 to 2016. Compared to CHIRPS the downscaled precipitation shows a higher precipitation mainly in mountainous regions such as the Andes, the Himalayas, the Alps, the Western Ghats, and Central Papua (Fig. 5). Lower precipitation can be found in most of the South East Asian Archipelago, Eastern China, and parts of the Amazon basin. At low elevation terrain, the differences between the two datasets are less pronounced.

For the extreme temperatures there are overall differences with the downscaled data showing lower maximum temperatures compared to CRU TS v4.01 and higher minimum temperatures in low elevation terrain (Fig. 5). In mountainous terrain the downscaled data shows lower minimum temperatures than CRU TS v4.01. Maximum temperature is mainly higher in arid regions such as in the Atacama, the southern Arabian Peninsula, parts of the Sahara, western North America, and in parts of Central Asia. We acknowledge that these differences might both come from errors in the CHELSA dataset, the CRU TS v4.01, or the CHIRPS data. Although the CRU data is based on observation, meteorological stations also exhibit an error by themselves. Meteorological stations for example tend to be sparse or non-existent and topographically biased towards low elevations in mountainous regions[66–68].

Precipitation gauges can underestimate snowfall by up to 90% due to wind-induced undercatch[63,69–71]. Both the downscaled CHELSA as well as the CHIRPS dataset have such biases that effect mainly high elevation and high latitude regions[63]. The high temperature differences between CRU and CHELSA_CMIP5 ts might come partly from the low data density in regions such as Amazonia, the Sahara, the Congo Basin, or Greenland. In many of these regions, temperature data is interpolated over large distances, leading to errors in the observational dataset. However, there are still differences in the temperature estimates compared to the GHCN data. These differences possibly stem from the simple change factor method applied. A more sophisticated downscaling such as quantile mapping[35,36,38] would possibly improve the temperature downscaling further.

In conclusion, the applied downscaling algorithm provided a better representation for precipitation and temperature estimates and offers a higher resolution estimate of GCM output for impact studies which require high horizontal resolutions.

## Usage Notes

A few caveats need to be considered with respect to the use of the data. The monthly data has now explicit correction for the frequency of e.g. extreme events which would only manifest on a daily resolution, therefore we do not recommend its use for studies involving climatic extremes. Temporal variability also comes solely from the respective driving GCM, and therefore temporal variabilities in either temperature or precipitation (e.g a low-frequency variability in ENSO) are equal to those of the respective GCM. Additionally, there is no correction for systematic gauge undercatch[63], so we caution against application of the data to calculate e.g. river discharge, hydrological modelling, or water resources assessments without using an additional bias correction for gauge undercatch. Such a bias-correction can however be applied by using the bias correction layers located here: http://www.gloh2o.org/pbcor/ which are compatible with CHELSA$_{CMIP5}$ts.

All CHELSA$_{CMIP5}$ts products are in a geographic coordinate system referenced to the WGS 84 horizontal datum, with the horizontal coordinates expressed in decimal degrees. The CHELSA layer extents (minimum and maximum latitude and longitude) are a result of the coordinate system inherited from the 1-arc-second GMTED2010 data which itself inherited the grid extent from the 1-arc-second SRTM data.

Note that because of the pixel center referencing of the input GMTED2010 data the full extent of each CHELSA grid as defined by the outside edges of the pixels differs from an integer value of latitude or longitude by 0.000138888888 degree (or 1/2 arc-second). Users of products based on the legacy GTOPO30 product should note that the coordinate referencing of CHELSA (and GMTED2010) and GTOPO30 are not the same. In GTOPO30, the integer lines of latitude and longitude fall directly on the edges of a 30-arc-second pixel. Thus, when overlaying CHELSA with products based on GTOPO30 a slight shift of 1/2 arc-second will be observed between the edges of corresponding 30-arc-second pixels.

The netCDF-4 formatted output files may be read with any netCDF tools that are CF-1.4 compliant.

The data are feely available under the Creative Commons Licence: CC BY.

## Code availability

The codes used are written in C++ and R and are included in SAGA GIS Version 6.1.0, freely available at www.saga-gis.org under the GNU public license including the necessary source codes. Calculations were done in SAGA Version 6.1.0.

## Acknowledgements

We want to thank Gian-Kasper Plattner for valuable comments on the selection of GCMs. DNK & NEZ acknowledges funding from the WSL internal project exCHELSA.

## Author contributions

D.N.K., G.D. and N.E.Z. developed the code, G.D. developed the scripts for cluster parallelisation of the code, D.R.S. handled the data curation, D.N.K. and D.R.S. validated the data, D.N.K. and N.E.Z. led the writing, and all authors contributed significant to revisions.

## Competing interests

The authors declare no competing interests.

## Additional information

**Supplementary information** is available for this paper at https://doi.org/10.1038/s41597-020-00587-y.

**Correspondence** and requests for materials should be addressed to D.N.K.

**Reprints and permissions information** is available at www.nature.com/reprints.

**Publisher's note** Springer Nature remains neutral with regard to jurisdictional claims in published maps and institutional affiliations.